\documentclass[preprint,12pt]{elsarticle}

\usepackage{graphics}

\usepackage{amssymb}

\begin{document}

\begin{frontmatter}


\title{Vacuum Solution of a Linear Red-Shift Based Correction in $f(R)$ Gravity
}
\author{Solmaz Asgari
}
\address{Department of Science,
Islamic Azad University, Abhar Branch, P.O.Box 22, Abhar, Iran} 
\ead{asgari\underline{ }s@iau$-$abhar.ac.ir}
\author{Reza Saffari\corref{cor1}
}
\address{Department of Physics,
University of Guilan, P. O. Box: 41335-1914, Rasht, Iran} 
\ead{rsk@guilan.ac.ir}
\cortext[cor1]{Corresponding author}



\begin{abstract}
In this paper we have considered a red-shift based linear correction
in derivative of action in the context of vacuum $f(R)$ gravity.
Here we have found out that the linear correction may describe the
late time acceleration which is appeared by SNeIa with no need of
dark energy. Also we have calculated the asymptotic action for the
desired correction. The value of all solutions may reduce to de'
Sitter universe in the absence of correction term.
\end{abstract}


\end{frontmatter}


\section{Introduction}
\label{int}

The recent data coming from the luminosity distance of SuperNovae Ia
(SNeIa) \cite{sn1a}, wide surveys on galaxies \cite{cole} and the
anisotropy of cosmic microwave background radiation \cite{cmb}
suggest that the Universe is undergoing an accelerating expansion.
Large Scale Structure formation \cite{lss}, Baryon Oscillations
\cite{bao} and Weak Lensing \cite{weak} also suggest such an
accelerated expansion of the Universe. Actually, identifying the
cause of this late time acceleration is one of the most challenging
problems of modern cosmology. Several approaches being responsible
for this expansion, have been proposed in the literature. A positive
cosmological constant can lead to accelerated expansion of the
universe but it is plagued by the fine tuning problem \cite{lcdm}.
The cosmological constant may be interpreted either geometrically as
modifying the left hand side of Einstein's equation or as a
kinematic term on the right hand side with the equation of state
parameter $w=-1$. Another approach can further be generalized by
considering a source term with an equation of state parameter
$w<-1/3$. Such kinds of source terms have collectively come to be
known as Dark Energy. Various scalar field models of dark energy
have been considered in literature \cite{scalar1}. All the dark
energy based theories assume that the observed acceleration is the
outcome of the action of a still unknown ingredient added to the
cosmic pie. In terms of the Einstein equations, $G_{\mu \nu} = \chi
T_{\mu \nu}$, such models are simply modifying the right hand side
including in the stress--energy tensor with something more than the
usual matter and radiation components.

On the other hand, as a radically different approach, one can also
try to leave unchanged the source side, but rather than modifying
the left hand side of Einstein field equations. In a sense, one is
therefore interpreting cosmic acceleration as a first signal of the
breakdown of the laws of physics as described by the standard
General Relativity (GR). Extending GR, not simply given its positive
results, opens the way to a large class of alternative theories of
gravity ranging from extra dimensions \cite{DGP} to non-minimally
coupled scalar fields \cite{stnoi}. In particular, we will be
interested here in fourth order theories \cite{fognoi} based on
replacing the scalar curvature $R$ in the Hilbert--Einstein action
with a generic analytic function $f(R)$ which should be
reconstructed starting from data and physically motivated issues.
Also referred to as $f(R)$ gravity, these models have been shown to
be able to fit both the cosmological data and Solar System
constraints in several physically interesting cases \cite{Hu}. These
theories are also referred to as 'extended theories of gravity',
since they naturally generalize General Relativity. It has been
predicted that the universe might have been appeared from an
inflationary phase in the past. It is also believed that the present
universe is passing through a phase of the cosmic acceleration.
Vacuum solutions of $f(R)$ gravity theories are one of interesting
subjects which are obtained for constant Ricci scalar
\cite{Nojiri,Nojiri1,Nojiri2,Nojiri3}, while it is possible to
derive non constant curvature scalar vacuum solutions.

In this paper we would like to note that $f(R)$ gravity theory is a
powerful approach to describe dynamical behavior of the Universe via
an unusual approach. Actually we consider vacuum solutions of $f(R)$
gravity. But there is a difference with other vacuum solutions. This
way we do not assume constant scalar curvature to obtain vacuum
solution. Vacuum solutions of modified $f(R)$ gravity would like to
explain the late time phase transition of cosmological parameters
like deceleration parameter without the need for dark companion of
the Universe, just by pure geometry.

The pioneering works on reconstruction of modified action through
inverse method are done by the authors of Ref.
\cite{Nojiri,Nojiri1,Nojiri2,Nojiri3}. They developed a general
scheme for cosmological reconstruction of modified $f(R)$ gravity in
terms of e-folding (or red-shift) without using auxiliary scalar in
intermediate calculations. Using this method, it is possible to
construct the specific modified gravity which contains any requested
FRW cosmology. The number of $f(R)$ gravity examples is used where
the following background evolutions may be realized: $\Lambda CDM$
epoch, deceleration with subsequent transition to effective phantom
superacceleration leading to Big Rip singularity, deceleration with
transition to transit phantom phase without future singularity,
oscillating universe. It is important that all these cosmologies may
be realized only by modified gravity without the use of any dark
components. In this essay, we try to reconstruct an appropriate
action for the modified gravity through the semi-inverse solution
method. We do not assume any FRW cosmology to reconstruct its
related $f(R)$ action. Our starting point is some modification in
deriving from a generic action which is depended on red-shift.

In section II, we have a briefer review of modified field equations.
In section III, we introduce our model and its results in field
solutions. In section IV, we study the evolution of deceleration
parameter under considered model and its related dark energy
Equation of State (EoS). In section V, we calculate an approximated
value for correction parameter which is in accordance with SNeIa
data from observational constraints. In section VI, we try to
reconstruct the original action which may produce our desired
corrections and in section VII we examine the local tests for the
obtained action. Section VIII is conclusion of this paper.

\section{Modified field equations}
\label{mfe} The action of modified theory of gravity is given by
\begin{equation}
S=\int d^4x\sqrt{-g}
\bigg[\frac{1}{2\kappa}f(R)+L_m\bigg],\label{action}
\end{equation}
where $L_m$ is the matter action such as radiation, baryonic matter,
dark matter and so on which we do not consider them in field
equation. In this essay, we consider the flat Friedmann Robertson
Walker, (FRW) background, so that the gravitational field equations
for $f(R)$ modified gravity are provided by the following form
\begin{equation}
-3\frac{\ddot{a}}{a}f'+3\frac{\dot{a}}{a}\dot{R}f''
+\frac12f=0,\label{modfrw1}
\end{equation}
\begin{equation}
[\frac{\ddot{a}}{a}+2\frac{\dot{a}^2}{a^2}]f'
-2\frac{\dot{a}}{a}\dot{R}f''-\dot{R}^2f'''
-\ddot{R}f''-\frac12f=0.\label{modfrw2}
\end{equation}
where the overdot denotes a derivative with respect to $t$, {\it and
the prime denotes a derivative with respect to $R$,} $a(t)$ is the
scale factor and $H=\dot{a}(t)/a(t)$ is the Hubble parameter.
Eliminating $f$ between Eqs. (\ref{modfrw1}) and (\ref{modfrw2})
results:
\begin{equation}
-2[\frac{\ddot a}{a}-(\frac{\dot a}{a})^2]f'+\frac{\dot a }{a}\dot
Rf''-\ddot Rf''-\dot R^2f'''=0 .\label{elim}
\end{equation}
Which can be changed in the form of:
\begin{equation}
2\dot HF-H\dot F+\ddot F=0,\label{diff1}
\end{equation}
where $F=df/dR$. Eq. (\ref{diff1}) is a second order differential
equation of $F$ with respect to time, in which both of $F$ and $H$
are undefined. The usual method to solve Eq. (\ref{diff1}) is based
on definition of $f(R)$.

Changing the variable of the above equation from $t$ to a new
variable like the number of e-folding, $N$ was done in
\cite{Nojiri1,Nojiri2,Nojiri3}. The variable $N$ is related to the
redshift, $z$ by $e^{-N}=1+z$. They solve the cosmological dynamic
equation by definition of Hubble parameter as a function of $N$ in a
general form. Then they rewrite the equation by redefinition of the
variable from the number of e-folding to the Ricci scalar and solve
it with respect to the Ricci scalar. Thus they could demonstrate
that modified $f(R)$ gravity may describe the $\Lambda CDM$ epoch
without any need for introducing the effective cosmological
constant, non-phantom matter and phantom matter. In this paper we
would like to replace the variable of Eq. (\ref{diff1}) by
red-shift, $z$, directly to solve the dynamical equation for a
specific red-shift depended action.

Each redshift, $z$ has an associated cosmic time $t$ (the time when
objects are observed with redshift $z$ emitting light), so we can
replace all the differentials with respect to $t$ by $z$ via:
\begin{eqnarray}
\frac{d}{dt}&=&\frac{da}{dt}\frac{dz}{da}\frac{d}{dz},\nonumber\\
&=&-(1+z)H(z)\frac{d}{dz},\label{rep1}
\end{eqnarray}
where we use $1+z={a_0}/{a}$, and we consider $a_0=1$, in the
present time. Now, we can replace the variable of Eq. (\ref{diff1})
from $t$ to $z$ by using Eq. (\ref{rep1}), and we obtain a first
order differential equation for $H^2$ with respect to $z$ as:
\begin{equation}
\frac{d}{dz}H(z)^2=P(z)H(z)^2,\label{diff3}
\end{equation}
where $P(z)$ is a function with respect to $z$, which depends on the
definition of $F$ with respect to $z$ as:
\begin{equation}
P(z)=\frac{2(1+z)(d^2F/dz^2)+4(dF/dz)}{2F-(1+z)(dF/dz)}.\label{pz}
\end{equation}
Now we may solve Hubble parameter that depends on the definition of
$F(z)$. This may happen by analytical calculations or numerical
approaches which is based on selection of $F(z)$. Since there could
be many choices to select the function of $F(z)$, we have decided to
add a linear correction term to General Relativistic limit of $F(R)$
value, which is discussed in the following section.

\section{Model selection and its solution}
\label{ms} Since we do not know the determined function of $F(z)$,
would like to consider a linear correction of red-shift as
\begin{eqnarray}
F(z)=F_0(1+\alpha z),\label{teyf}
\end{eqnarray}
in which for $\alpha=0$ we have $F(z)=F_0$. In this case Eq.
(\ref{diff3}) has a solution such as $H(z)^2=H_0^2$ in which $H_0$
is a constant. This case will reproduce de' Sitter solution of
vacuum Universe that is expanded with a constant velocity. Since the
value of $H_0$ is independent of $F_0$ we may fix it to $F_0=1$ to
recover GR or transfer its effect to gravitational constant by
normalization. For $\alpha\neq 0$ we have our linear correction. In
this case Eq. (\ref{diff3}) has a solution such as
\begin{eqnarray}
H(z)^2=H_\alpha^2\big[1-\frac12\alpha(1-z)\big]^4,\label{m1h}
\end{eqnarray}
where $H_\alpha^2=H_0^2/(1-\alpha/2)^4$. It is clear that it may
reduce to de' Sitter case in the absence of $\alpha$.

\section{Cosmic dynamics and EoS}
\label{cd}

\begin{figure}
\includegraphics{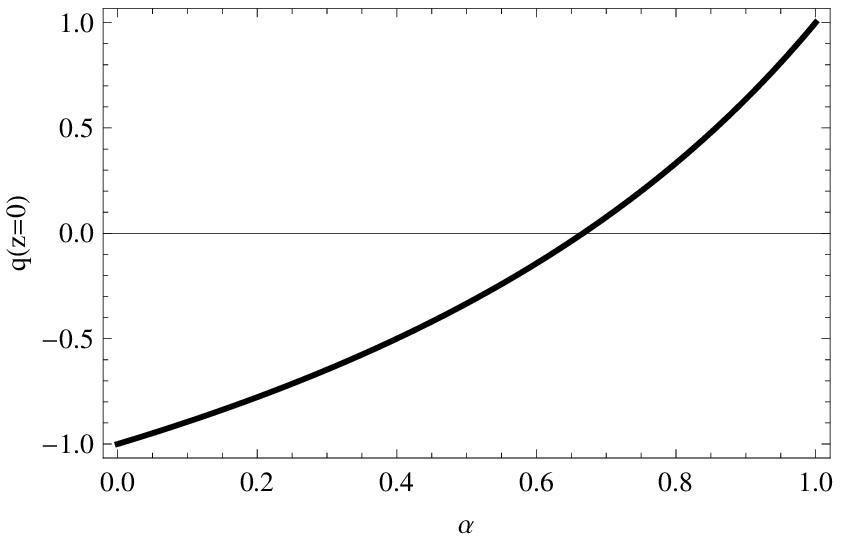}
\caption{\label{q0}$q(z=0)$ with respect to $\alpha$.}
\end{figure}

\begin{figure}
\includegraphics{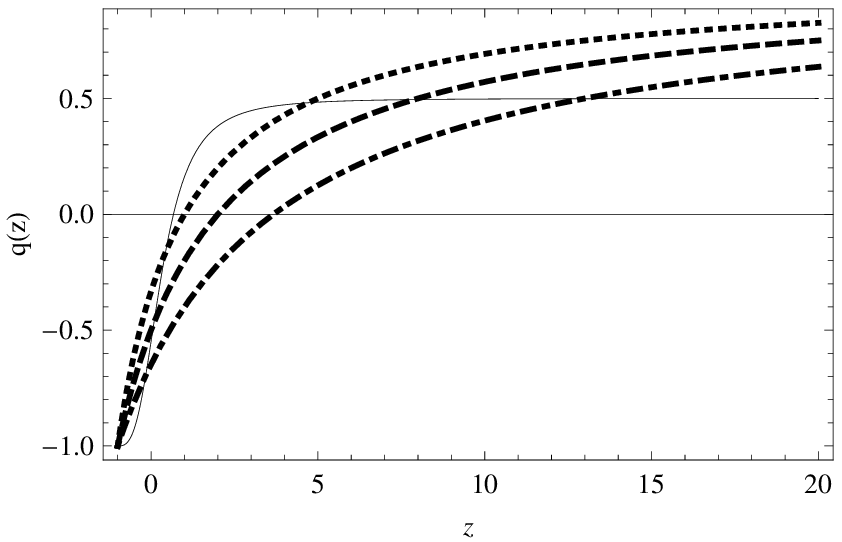}
\caption{\label{fqofz}$q(z=0)$ with respect to $\alpha$.}
\end{figure}

Deceleration parameter, $q$ in cosmology is a dimensionless measure
of the cosmic acceleration of the expansion of space in a FRW
universe. It is defined by:
\begin{equation}
q=-\frac{\ddot aa}{\dot a^2}=-1-\frac{\dot H}{H^2}.\label{q}
\end{equation}
here we change the variable of Eq. (\ref{q}) from $t$ to $z$ then we
will have evolution of deceleration parameter with respect to
redshift as:
\begin{eqnarray}
q(z)&=&-1+\frac{1+z}{2H(z)^2}\frac{dH(z)^2}{dz},\label{qz1}\\
&=&-1+\frac{1+z}{2}P(z),\label{qz2}
\end{eqnarray}
which only depends on $P(z)$ for vacuum solution. Since we would
like to study the first order correction of $F(z)$ we put the
modified model in Eq. (\ref{qz2}) to obtain
\begin{equation}
q(z)=-1+\frac{2\alpha(1+z)}{2-\alpha(1-z)}.\label{qofz}
\end{equation}
Transition point from deceleration to acceleration phase ($q(z)=0$)
obtains as
\begin{equation}
z_T=\frac{2 - 3 \alpha}{\alpha},\label{tp}
\end{equation}
which is a constraint for our correction parameter as $0<\alpha<2/3$
to have a transition point in recent positive red-shifts, but it
should not be close to zero because this value will shoot the
transition to high red-shifts. On the other hand as it is shown in
Fig. (\ref{q0}) for all the values of $\alpha<2/3$ universe is under
acceleration.
Also evolution of deceleration parameter from high red-shifts to the
future is shown in Fig. (\ref{fqofz}). Another result of Eq.
(\ref{qofz}) is for large red-shifts, $q(z)\rightarrow 1$ which is
the value of deceleration parameter in radiation era.

\begin{figure}
\includegraphics{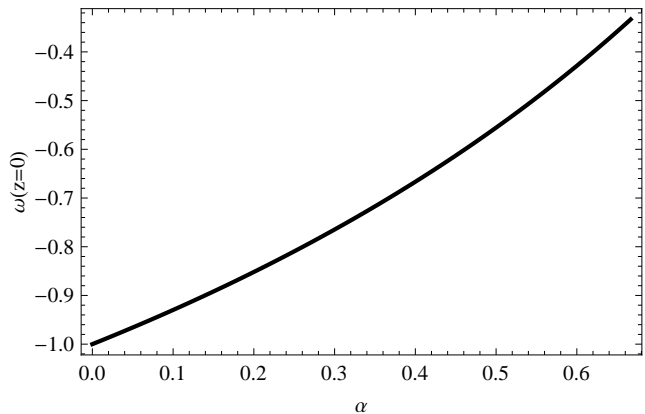}
\caption{\label{omofa}$\omega(z=0)$ with respect to $\alpha$.}
\end{figure}

\begin{figure}
\includegraphics{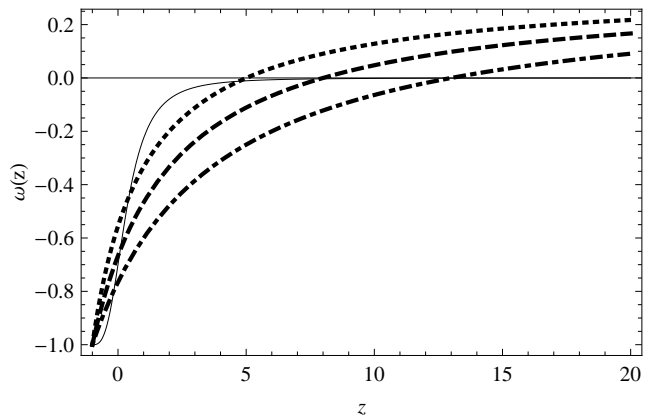}
\caption{\label{omofz}Evolution of $\omega(z)$ for $\Lambda CDM$
(solid line) with $\Omega_m=0.3$ and $\Omega_\Lambda=0.7$, and for
our model with correction parameters $\alpha=0.5$ (dotted line),
$\alpha=0.4$ (dashed line) and $\alpha=0.3$ (dashed-dotted line).}
\end{figure}

The equivalent form of dark energy Equation of State (EoS) which
corresponds to selected $F(z)$ is obtained as
\begin{eqnarray}
\omega(z)&=&-1+\frac{1+z}{3H(z)^2}\frac{dH(z)^2}{dz},\label{state}\\
&=&-1+\frac{4\alpha(1+z)}{3[2-\alpha(1-z)]},\label{omofze}
\end{eqnarray}
which is reduced to de' Sitter EoS that is equal to $\omega=-1$ when
$\alpha=0$. Fig. (\ref{omofa}) shows the range of the predicted
value of EoS with respect to acceptable values of $\alpha$.
Also evolution of EoS from high red-shifts to low red-shifts is
shown in Fig. (\ref{omofz}). On the other hand Eq. (\ref{omofze})
shows that for red-shifts that are large enough,
$\omega(z)\rightarrow 1/3$ which is EoS in radiation era, while we
have solved modified Friedmann equations for vacuum.

\section{Supernovae constraint}
\label{cd}

\begin{figure}
\includegraphics{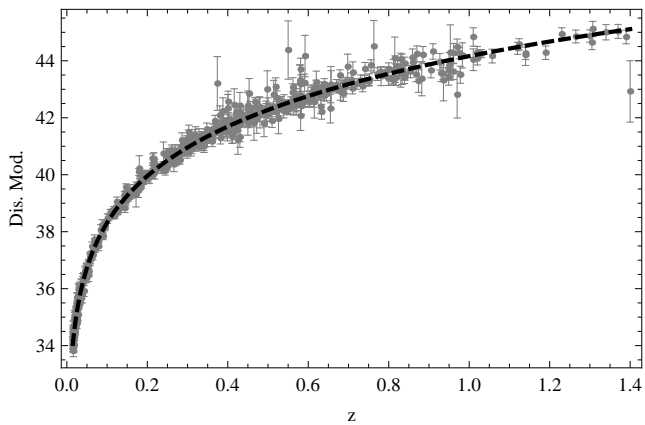}
\caption{\label{dismod}Accordance between Union2 SNeIa data and late
time linear correction of our model for $\alpha\approx0.418$ and
$h\approx0.697$.}
\end{figure}

In the following we would like to constraint our model parameters in
comparison with observational data sets. Another main late time
cosmological constraint which is tested for the above simple example
is SNeIa distance module with respect to redshift constraint. In
this comparison we used the Union2 data set \cite{union2} which
provide 557 SNeIa specifications. According to the value of
numerical results which was gotten in comparison with SNeIa data set
from usual $\chi^2$ algorithm, we discover that for $\alpha=0.418$,
$h=0.697$ where $h$ is the reduced Hubble parameter and
$\chi^2=545.329$, so the ratio of chi square error to the number of
freedom is about $0.983$. The result of this calculation is shown in
Fig. (\ref{dismod}).

\section{Reconstruction of $f(R)$ action}
\label{cd}

\begin{figure}
\includegraphics{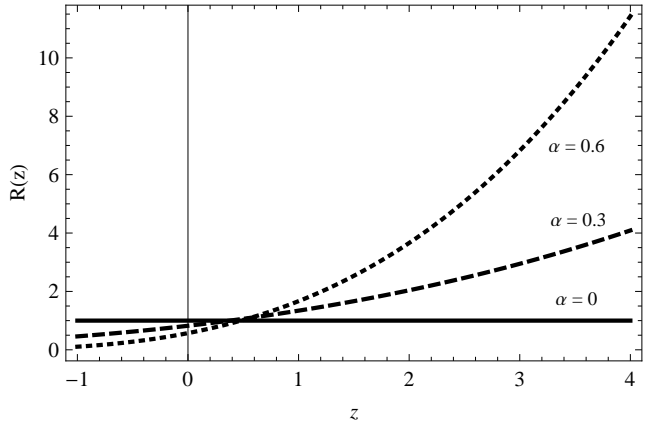}
\caption{\label{rofz}Evolution of $R/R_0$ with respect to $z$ for
three values of correction parameter. The crossing points are not
unique.}
\end{figure}

Specification of modified action is one of interesting subjects in
the story of $f(R)$ gravity. Here we may reconstruct the action via
a semi inverse solution approach. The Ricci scalar is defined as
\begin{equation}
R=6(\dot H+2H^2),\label{ricci1}
\end{equation}
and if we change the variable of this equation from time to
red-shift we have
\begin{equation}
R(z)=3[4H(z)^2-(1+z)\frac{dH(z)^2}{dz}].\label{ricci2}
\end{equation}
The Ricci scalar for the Hubble parameter of Eq. (\ref{m1h}) is
obtained from Eq. (\ref{ricci2}) as
\begin{equation}
R(z)=R_\alpha\bigg[1-\frac12\alpha(1-z)\bigg]^3,\label{sricci2}
\end{equation}
where
\begin{equation}
R_\alpha=\frac{12(1-\alpha)H_0^2}{(1-\alpha/2)^4},\label{Ra}
\end{equation}
which will be reduced to another constant in $\alpha=0$ and we may
name it the present value of Ricci scalar $R_0=12H_0^2$. Changes of
modified Ricci scalar is shown in Fig. (\ref{rofz}). If we suppose
the value of cosmological constant, $\Lambda=3H_0^2$ then we have a
simplified relation between the present value of curvature scalar
and cosmological constant as $R_0=4\Lambda$. We may use these
equalities to find behavior of the modified solutions in the
constraint regions.

\begin{figure}
\includegraphics{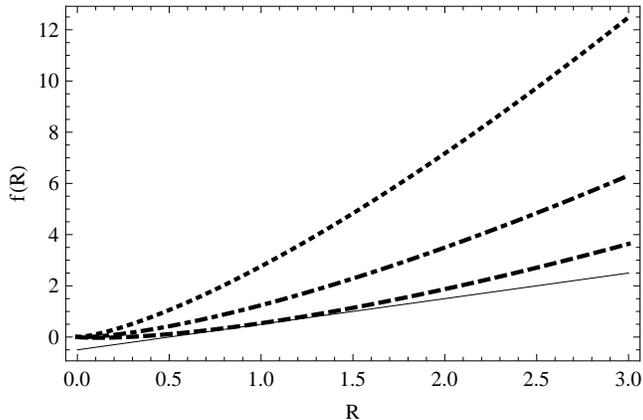}
\caption{\label{fofRofR}Change of $f(R)$ with respect to $R$ for
three values of correction parameter, $\alpha=0$ in dashed line,
$\alpha=0.3$ in dot-dashed line and $\alpha=0.6$ in dotted line.
Solid line shows variation of GR action with cosmological constant
which is de' Sitter action.}
\end{figure}
One of the simple solutions to achieve the desired action is
eliminating of $z$ between $F(z)$ and $R(z)$ which leads us to reach
the equation
\begin{equation}
F(R)=2(\frac{R}{R_\alpha})^{1/3}+\alpha-1.\label{FR1}
\end{equation}
It is clear that, the derivative of action is different from GR plus
cosmological constant or de' Sitter model. But in the case of
$\alpha=0$ or $R_\alpha=R_0$, we would like to recover de' Sitter
solution. Then by using Eq. (\ref{sricci2}) we obtain $R=R_0$ and
then $F(R)=1$ which is equal to the GR or de'Sitter model. Because
GR action differs from de'Sitter action only in a constant term.

Now we can calculate $f(R)$ by integrating of Eq. (\ref{FR1}) with
respect to $R$ as
\begin{equation}
f(R)=R\big[\frac32(\frac{R}{R_\alpha})^{1/3}+\alpha-1\big]+C,\label{fR1}
\end{equation}
where $C$ is constant of integration. Since we would like to recover
de' Sitter model in the absence of correction term, $\alpha$ we must
find the true constraint on the constant of integration. we can
determine the constraint by comparing the value of de' Sitter action
with the action of Eq. (\ref{fR1}) when $\alpha =0$ or $R=R_0$. The
standard form of de' Sitter action is $f_d(R)=R-2\Lambda$. Since the
scalar curvature of de' Sitter universe is a constant, then we can
determine the value of $f_d(R)$ in the presence of constant
curvature. If we solve the standard algebraic equation of $f(R)$
gravity as $2f_d(R)-RF_d(R)=0$ we can find $R_0=4\Lambda$. Then the
value of de' Sitter action in the case of $R=R_0$ is
$f_d(R_0=4\Lambda)=2\Lambda$ which is equal to $R_0/2$.

For the case of our obtained action from Eq. (\ref{fR1}), we have
$f(R=R_0)=R_0/2+C$ when $\alpha=0$. Since we would like to have the
same solution for de' Sitter model and our obtained action in
present of $R=R_0$ we force constant of integration, $C$ to be equal
to zero.

Therefore the obtained action in its constrained form would be
\begin{equation}
f(R)=R\big[\frac32(\frac{R}{R_\alpha})^{1/3}+\alpha-1\big],\label{fR2}
\end{equation}
which is different from de' Sitter action even in the limit of
$\alpha=0$, but it behaves like $f_d(R)=R-2\Lambda$ in the region of
$R\simeq R_0$ as shown in Fig. (\ref{fofRofR}). In the following of
this section we discuss one special case of parameter $\alpha$ to
see how the reconstructed action is related to the de' Sitter
action.
\subsection{$\alpha \ll 1$}
In this case the constant $R_\alpha$ is reduced to $R_\alpha\sim
R_0(1+\alpha)$. There are two subcases in this limit. The first one
is low red-shift approximation. In this case Ricci scalar in Eq.
(\ref{sricci2}) is reduced to $R(z)\sim R_0(1-\alpha/2)=const.$,
which is as same as the case of constant expansion rate or de'
Sitter universe, but reconstruction of $F(R)$ is impossible via our
approach. The second one is high red-shift approximation. In this
case Ricci scalar is reduced to $R(z)\sim R_0(1+\frac32\alpha z)$.
Then the derivative of action is obtained in the form of
\begin{equation}
F(R)=\frac13(1+2\frac{R}{R_0}).\label{FR100}
\end{equation}
Integration of the above equation with respect to $R$ obtains the
desired action as
\begin{equation}
f(R)=\frac13(R+\frac{R^2}{R_0})+C,\label{fR100}
\end{equation}
where $C$ is the constant of integration which plays the role of
effective cosmological constant. The factor of $1/3$ may transform
the gravitational constant and translate $G_N$ to $G_{eff}$. In this
case the form of the above action is reduced to de' Sitter action
when $R\rightarrow 0$ and effective cosmological constant defines as
$\Lambda_{eff}=3C$.


\section{Local tests}
\label{lt}

Following \cite{Chiba} here we introduce the auxiliary field $A$ to
rewrite geometric part of the action (\ref{action}) in the following
form:
\begin{equation}
S=\frac{1}{2\kappa}\int d^4 x \sqrt{-g} \left\{f'(A)\left(R-A\right)
+ f(A)\right\}.\label{JGRG21}
\end{equation}
By the variation over $A$, one obtains $A=R$. Substituting $A=R$
into the action (\ref{JGRG21}), one can reproduce the action in
(\ref{action}). Furthermore, we rescale the metric in the following
way (conformal transformation):
\begin{equation} \label{JGRG22}
g_{\mu\nu}\to {\rm e}^\sigma g_{\mu\nu}\ ,\quad \sigma =-\ln f'(A).
\end{equation}
Hence, the Einstein frame action is obtained:
\begin{eqnarray}
\label{JGRG23} S_E &=& \frac{1}{2\kappa}\int d^4 x \sqrt{-g} \left(
R - \frac{3}{2}g^{\rho\sigma}
\partial_\rho \sigma \partial_\sigma \sigma - V(\sigma)\right), \\
\nonumber V(\sigma) &=& {\rm e}^\sigma g\left({\rm
e}^{-\sigma}\right)
 - {\rm e}^{2\sigma} f\left(g\left({\rm e}^{-\sigma}\right)\right) = \frac{A}{f'(A)} -
 \frac{f(A)}{f'(A)^2}.
\end{eqnarray}
 Here $g\left({\rm e}^{-\sigma}\right)$ is given by solving the equation
$\sigma = \ln f'(A)$ as $A=g\left({\rm e}^{-\sigma}\right)$. Due to
the scale transformation (\ref{JGRG22}), a coupling of the scalar
field $\sigma$ with usual matter appears. The mass of $\sigma$ is
given by
\begin{equation} \label{JGRG24}
m_\sigma^2 \equiv \frac{1}{2}\frac{d^2 V(\sigma)}{d\sigma^2}
=\frac{1}{2}\left\{\frac{A}{f'(A)} -
\frac{4f(A)}{\left(f'(A)\right)^2} + \frac{1}{f''(A)}\right\}\ .
\end{equation}
Unless $m_\sigma$ is very  large, there  appears large correction to
the Newton law. Naively, one expects the order of the mass
$m_\sigma$ to be that of the Hubble rate, that is, $m_\sigma \sim
H_0 \sim 10^{-33}\,{\rm eV}$, which is very light and the correction
could be very large, which is claimed in \cite{Chiba}.

We should note, however, that the mass $m_\sigma$  depends on the
detailed form of $F(R)$ in general \cite{NO}. Moreover, the mass
$m_\sigma$  depends on the curvature. The curvature on the earth
$R_{\rm earth}$ is much larger than the average curvature $R_{\rm
solar}$ in the solar system and $R_{\rm solar}$ is also much larger
than the average curvature in the unverse, whose order is given by
the square of the Hubble rate $H^2$, that is, $R_{\rm earth} \gg
R_{\rm solar} \gg H^2$. Then if the mass becomes large when the
curvature is large, the correction to the Newton law could be small.
Such mechanism is called the Chameleon mechanism and proposed for
the scalar-tensor theory in \cite{KW}. In the case of action
(\ref{fR2}), the mass $m_\sigma$ is given by
\begin{equation} \label{JGRG25}
m_\sigma^2 \sim \frac{1}{4}R_{\alpha}^{1/3}A^{2/3},
\end{equation}
where $R_{\alpha}$ is defined in Eq. (\ref{Ra}). The order of
$R_\alpha$ is about $R_\alpha \sim 12H_0^2 \sim 10^{-65}\,{\rm
eV}^2$ for the value of $0<\alpha<2/3$. Then in solar system, where
$R\sim 10^{-61}\,{\rm eV}^2$, the mass is given by $m_\sigma^2 \sim
10^{-31}\,{\rm eV}$ and in the air on the earth, where $R \sim
10^{-50}\,{\rm eV}^2$, $m_\sigma \sim 10^{-28}\,{\rm eV}$. The order
of the radius of the earth is $10^7\,{\rm m} \sim
\left(10^{-14}\,{\rm eV}\right)^{-1}$, $m_\sigma \sim 10^{-13}\,{\rm
eV}$. Therefore the scalar field $\sigma$ is very light and the
correction to the Newton law is observable. Here we should note that
the obtained action (\ref{fR2}) strongly depends on the order of
red-shift based correction terms, strongly. Then one can consider
local tests under influence of higher order corrections in the
derivative of action.

\section{Conclusions}
In this paper we have examined a first order red-shift based
correction on derivative of general action with respect to Ricci
scalar as a starting point of modified gravity in $f(R)$ theory.
Also we have shown that this correction may operate as an
alternative for dark energy via its ability in according with
distance modulus of Union2 data set of SNeIa. There is an
interesting behavior in the limit of low curvature regions for the
obtained modified action. As it is clear in the region of
$R\rightarrow 0$ of Fig. (\ref{fofRofR}), the modified action tends
to the value of scalar curvature, $f(R)\rightarrow R$ for all of
values of correction term, which is pure General Relativity action
and in the limit of $R\rightarrow R_0$ it will be close to
$f(R)\rightarrow R-2\Lambda$ in the case of $\alpha=0$. However the
mass of equivalent scalar field of the obtained action is not heavy
enough to evade local tests of the theory, the behavior of other
spacetime solutions such as spherically symmetric space is
considerable. As it is shown in this paper, solution of modified
equation in the absence of baryonic matter and cosmic cold dark
matter could get better results in comparison with $\Lambda CDM$ in
the case of expansion rate of universe. In this manner the $f(R)$
may play as an alternative for cosmic dark matter. But $f(R)$
theories can also play a major role at astrophysical scales. In
fact, modifying the gravitational Lagrangian affects the
gravitational potential in the low energy limit. A corrected
gravitational potential could offer the possibility to fit galaxy
rotation curves without any need of huge amounts of dark matter,
which is considerable for the obtained action (\ref{fR2}). On the
other hand behavior of high red-shifts or early time cosmology of
the action is considerable, because the power law terms may be
considered to explain inflationary behavior of early universe. Of
course the main correction of this approach comes from selection of
$F(z)=1+\alpha z$ which is a toy model that could to explain late
time inflationary behavior of the universe without dark energy. This
approach is based on general behavior scalar curvature, not as a
constant. Also it will explain radiation dominated era values of
parameters such as deceleration parameter and EoS which is in
accordance with a universe including radiation density, while the
modified field equations have been solved for empty universe. Our
main scope from this paper is introducing an approach that may
proposed as an alternative for dark energy. So we can demonstrate if
the obtained action is viable for other scales of universe or not?
Finally we propose higher order red-shift based corrections to have
more accurate behavior of modified solutions which is in progress.

\section{Acknowledgement}
We would like to thank anonymous referees for useful comments. Also
we would like to thank Sh. Faghihzadeh for his grammatical
corrections.





\bibliographystyle{model1a-num-names}
\bibliography{<your-bib-database>}

\begin{thebibliography}{00}


\bibitem{sn1a}
S. Perlmutter {\it et al.}, Astrophys. J. {\bf 517} (1999) 565
[arXiv:astro-ph/9812133]; A. G. Riess {\it et al.}, Astron. J.{\bf
116} 1009 (1998) [arXiv:astro-ph/9805201].

\bibitem{cole}
S. Cole {\it et al.}, Mon. Not. Roy. Astron. Soc. {\bf 362}, 505
(2005) [arXiv:astro-ph/0501174].

\bibitem{cmb}
D.N.Spergel {\it et al}., [WMAP collaboration], Astrophys. J. Suppl.
{\bf 170}, 377 (2007) [arXiv:astro-ph/0603449].

\bibitem{lss}
U. Seljak {\it et al.}, [SDSS collboration], Phys. Rev. D {\bf 71},
103515 (2005) [arXiv:astro-ph/0407372].

\bibitem{bao}
D. J. Eisenstein {\it et al.}, [SDSS collboration], Astrophys. J.
{\bf 633}, 560 (2005) [arXiv:astro-ph/0501171].

\bibitem{weak}
B. Jain and A. Taylor, Phys. Rev. Lett. {\bf 91}, 141302 (2003)
[arXiv:astro-ph/0306046].

\bibitem{lcdm}
E. J. Copeland, M. Sami and S. Tsujikawa, Int. J. Mod. Phys. D {\bf
15}, 1753 (2006); M. Sami, Curr. Sci. {\bf 97}, 887 (2009)
[arXiv:0904.3445]; V. Sahni and A. A. Starobinsky, Int. J. Mod.
Phys. D {\bf 9}, 373 (2000); T. Padmanabhan, Phys. Rep. {\bf 380},
235 (2003); E. V. Linder, Gen. Rel. Grav. {\bf 40}, 329 (2008)
[arXiv:0704.2064]; J. Frieman, M. Turner and D. Huterer, Ann. Rev.
Astron. Astrophys. {\bf 46}, 385 (2008) [arXiv:0803.0982]; R.
Caldwell and M. Kamionkowski, Ann. Rev. Nucl. Part. Sci. {\bf 59},
397 (2009) [arXiv:0903.0866]; A. Silvestri and M. Trodden, Rept.
Prog. Phys. {\bf 72}, 096901 (2009) [arXiv:0904.0024].

\bibitem{scalar1}
C. Armendariz-Picon, T. Damour, and V. Mukhanov, Phys. Lett. B {\bf
458}, 209 (1999); J. Garriga and V. F. Mukhanov, Phys. Lett. B {\bf
458}, 219 (1999); T. Chiba, T. Okabe, M. Yamaguchi, Phys. Rev. D
{\bf 62}, 023511 (2000); C. Armendariz-Picon, V. Mukhnov, and P. J.
Steinhardt, Phys. Rev. Lett {\bf 85}, 4438 (2000); C.
Armendariz-Picon, V. Mukhnov, and P. J. Steinhardt, Phys. Rev. D
{\bf 63}, 103510 (2001); T. Chiba, Phys. Rev. D {\bf 66}, 063514
(2002); L. P. Chimento and A. Feinstein, Mod. Phys. Lett. A {\bf
19}, 761 (2004); L. P. Chimento, Phys. Rev. D {\bf 69}, 123517
(2004); R. J. Scherrer, Phys. Rev. Lett. {\bf 93}, 011301 (2004); A.
Y. Kamenshchik, U. Moschella, and V. Pasquier, Phys. Lett. B {\bf
511}, 265 (2001); N. Bilic, G. B. Tupper, and R. D. Viollier, Phys.
Lett. B {\bf 535}, 17 (2002); M. C. Bento, O. Bertolami, and A. A.
Sen, Phys. Rev. D {\bf 66}, 043507 (2002); A. Dev, J. S. Alcaniz,
and D. Jain, Phys. Rev. D {\bf 67}, 023515 (2003); V. Gorini, A.
Kamenshchik and U. Moschella, Phys. Rev. D {\bf 67}, 063509 (2003);
R. Bean and O. Dore, Phys. Rev. D {\bf 68}, 23515 (2003); T.
Multamaki, M. Manera and E. Gaztanaga, Phys. Rev. D {bf 69}, 023004
(2004); A. A. Sen and R. J. Scherrer, Phys. Rev. D {\bf 72}, 063511
(2005).

\bibitem{DGP}
G.R. Dvali, G. Gabadadze and M. Porrati, Phys. Lett. B, {\bf 485},
208 (2000); G. R. Dvali, G. Gabadadze, M. Kolanovic and F. Nitti,
Phys. Rev. D, {\bf 64}, 084004 (2001); G. R. Dvali, G. Gabadadze, M.
Kolanovic and F. Nitti, Phys. Rev. D, {\bf 64}, 024031 (2002); A.
Lue, R. Scoccimarro and G. Starkman, Phys. Rev. D, {\bf 69}, 044005
(2004); A. Lue, R. Scoccimarro and G. Starkman, Phys. Rev. D, {\bf
69}, 124015 (2004).

\bibitem{stnoi}
P. Caresia, S. Matarrese and L. Moscardini, Astrophys. J., {\bf
605}, 21 (2004); V. Pettorino, C. Baccigalupi and G. Mangano, JCAP,
{\bf 0501}, 014 2005; M. Demianski, E. Piedipalumbo, C. Rubano and
C. Tortora, Astron. Astrophys., {\bf 454}, 55 (2006); S. Thakur, A.
A. Sen and T. R. Seshadri, Phys. Lett. B {\bf 696}, 309 (2011).

\bibitem{fognoi}
S. Capozziello, Int. J. Mod. Phys. D, {\bf 11}, 483 (2002); S.
Capozziello, V. F. Cardone, S. Carloni and A. Troisi, Int. J. Mod.
Phys. D, {\bf 12}, 1969 (2003); S. Capozziello, V. F. Cardone and A.
Trosi, Phys. Rev. D, {\bf 71}, 043503 (2005); S. Carloni, P. K. S.
Dunsby, S. Capozziello and A. Troisi, Class. Quant. Grav., {\bf 22},
4839 (2005); H. Kleinert and H. J. Schmidt, Gen. Rel. Grav.  {\bf
34}, 1295 (2002); S. Nojiri and S.D. Odintsov, Phys. Lett. B, {\bf
576}, 5 (2003); S. Nojiri and S. D. Odintsov, Mod. Phys. Lett. A,
{\bf 19}, 627 (2003); S. Nojiri and S. D. Odintsov, Phys. Rev. D,
{\bf 68}, 12352 (2003); S. M. Carroll, V. Duvvuri, M. Trodden and M.
Turner, Phys. Rev. D, {\bf 70}, 043528 (2004); G. Allemandi, A.
Borowiec and M. Francaviglia, Phys. Rev. D {\bf 70}, 103503 (2004);
S. Nojiri and S. D. Odintsov, Int. J. Geom. Meth. Mod. Phys. {\bf
4}, 115 (2007); S. Capozziello, M. Francaviglia, Gen. Rel. Grav.
{\bf 40}, 357 (2008); V. Faraoni, Rev. Mod. Phys. {\bf 82}, 451
(2010).

\bibitem{Hu}
W. Hu and I. Sawicki, Phys. Rev. D, {\bf 76}, 064004 (2007); A. A.
Starobinsky, JETP Lett., {\bf 86}, 157 (2007); S. A. Appleby and R.
A. Battye, Phys. Lett. B, {\bf 654}, 7 (2007); S. Nojiri and S. D.
Odintsov, Phys. Lett. B, {\bf 652}, 343 (2007); S. Tsujikawa, Phys.
Rev. D {\bf 77}, 023507 (2008); R. Hashemi and R. Saffari, Planet.
Space Sci. {\bf 59}, 338 (2011); S. Asgari and R. Saffari, Appl.
Phys. Res. {\bf 2}, 99 (2010); R. Saffari and S. Rahvar, Mod. Phys.
Lett A {\bf 24}, 305 (2009); R. Saffari and S. Rahvar, Phys. Rev. D
{\bf 77}, 104028 (2008); S. Baghram and S. Rahvar, Phys. Rev. D {\bf
80}, 124049 (2009).

\bibitem{Nojiri}
S. Nojiri and S. D. Odintsov, Phys. Rev. D {\bf 74}, 086005 (2006);
S. Nojiri and S. D. Odintsov, Phys. Lett. B {\bf 657}, 238 (2007);
G. Cognola, E. Elizade, S. Nojiri, S. D. Odintsov, L. Sebastiani,
and L. Zerbini, Phys. Rev. D {\bf 77}, 046009 (2008); E. Elizade, S.
Nojiri, S. D. Odintsov, L. Sebastiani, and L. Zerbini, Phys. Rev. D
{\bf 83}, 086006 (2011).

\bibitem{Nojiri1}
S. Nojiri and S. D. Odintsov, J. Phys. Conf. Ser. {\bf 66}, 012005.
(2007)

\bibitem{Nojiri2}
S. Nojiri and S. D. Odintsov and D. Saez-Gomez, [arXiv: 0908.1269]

\bibitem{Nojiri3}
S. Nojiri and S. D. Odintsov, [arXiv: 1011.0544].

\bibitem{Chiba}
Takeshi Chiba, Phys. Lett. B {\bf 575}, 1 (2003); S. Nojiri and S.
D. Odintsov, [arXiv: 0807.0685].

\bibitem{NO}
S. Nojiri and S. D. Odintsov, Phys. Rev. D {\bf 68}, 123512 (2003).

\bibitem{KW}
J. Khoury and A. Weltman, Phys. Rev. D {\bf 69}, 044026 (2004).

\bibitem{union2}
Amanullah, R., et al., Astrophys. J. {\bf 716} 712, (2010).



%
\end{thebibliography}



%

\end{document}